\documentclass[11pt,a4paper]{article}
\pdfoutput=1
\usepackage{jheppub}

\makeatletter
\def\@fpheader{\relax}
\makeatother

\usepackage{fancyhdr}
\usepackage{amsmath,amsthm,amssymb}
\usepackage{subcaption}
\usepackage{enumerate}
\usepackage{color}
\usepackage{graphicx}
\usepackage{caption}

\newcommand{\dd}{\mathrm{d}}

\def\beq{\begin{equation}}
\def\eeq{\end{equation}}
\def\bea{\begin{eqnarray}}
\def\eea{\end{eqnarray}}

\def\bwt{\begin{widetext}}
\def\ewt{\end{widetext}}

\newcommand{\tens}[1]{%
  \mathbin{\mathop{\bigotimes}\limits_{#1}}%
}

\begin{document}

\title{Continuum Modes of Nonlocal Field Theories}

\author{Mehdi Saravani}
\affiliation{School of Mathematical Sciences, University of Nottingham, University Park, Nottingham, UK}
\emailAdd{mehdi.saravani@nottingham.ac.uk}

\abstract{A class of nonlocal Lorentzian quantum field theories is introduced in \cite{Saravani:2015rva,Belenchia:2014fda}, where the d'Alembertian operator $\Box$ is replaced by a non-analytic function of the d'Alembertian, $f(\Box)$. This is inspired by the Causal Set program where such an evolution arises as the continuum limit of a wave equation on causal sets. The spectrum of these theories contains a continuum of massive excitations. This is  perhaps the most important feature which leads to distinct/interesting phenomenology. In this paper, we study properties of the continuum massive modes in depth. We derive the path integral formulation of these theories. Meanwhile, this derivation introduces a dual picture in terms of local fields which clearly shows how continuum massive modes of the nonlocal field interact. The dual picture, in principle, provides a path to extension beyond scalar fields and addressing the issue of renormalization.}

\maketitle

\section{Introduction}\label{introduction}


In this manuscript, we explore a class of nonlocal quantum field theories, proposed and studied in \cite{Saravani:2015rva,Belenchia:2014fda}. In these theories, the d'Alembertian operator $\Box$ is replaced by a nonlocal operator $f(\Box)$ where $f$ is a non-analytic function with a branch cut on real negative values. We refer to this class as non-analytic quantum field theories (NAQFT). 

NAQFT arise as the low energy limit of quantum field theories on Causal Set \cite{Benincasa:2010ac,Aslanbeigi:2014zva,Dowker:2013vba}, an approach to quantum gravity which replaces spacetime continuum with a discrete structure \cite{Bombelli:1987aa}. 
Lorentz invariance and nonlocality are two generic features of field theories on a manifold like causal set background. A residue of nonlocality persists in the low energy limit when the causal set is approximated by a continuum spacetime \cite{Sorkin:2007qi}.
This is an important motivation to study NAQFT, as they provide new posibilities for Causal Set phenomenology. 

Although these theories are nonlocal, the nonlocality is introduced in a way that keeps Lorentz invariance intact, i.e. nonlocality in space and time. Consequently, NAQFT also provide a framework to test Lorentzian nonlocality in nature. This is an important question on its own, regardless of the origin of nonlocality. Ref. \cite{Belenchia:2016sym,Alkofer:2016utc,Belenchia:2016zaa} investigated low energy phenomenology of these theories, focusing on deviations from local field theory counterparts due to nonlocal modifications.

In this paper, we are mainly concerned with another aspect of NAQFT. The spectrum of these theories contains a continuum of massive excitations, unlike local field theories where modes are confined to a mass shell. Due to the distinct behaviour of continuum massive modes, the authors in \cite{Saravani:2015rva} proposed them as a dark matter candidate. This was further investigated in \cite{Saravani:2016enc} in the context of cosmology.
Our goal here is to investigate the properties of continuum massive modes and introduce a more intuitive picture of NAQFT. In doing so, we come up with an equivalent description of NAQFT in terms of local fields. This description clearly shows how the continuum massive modes of the nonlocal field behave and provides a new interpretation of the previous results on NAQFT. With this goal in mind, let us review the free theory.

We use the following notation throughout the paper: spatial quantities are written as bold characters, e.g. $p=(p^0,\bf p)$ and $x=(x^0,{\bf x})$. Also, $p\cdot x=\eta_{\mu\nu}p^\mu x^\nu=-x^0p^0+\bf x\cdot \bf p$, where $\eta_{\mu\nu}$ is the Minkowski metric with mostly positive signature. 
\subsection{Free nonlocal field theory}\label{free_field}
Let us start by specifying the correlation functions of the nonlocal free scalar field theory \cite{Saravani:2015rva,Belenchia:2014fda}. For a free field, the one point and two point functions completely fix the theory. The nonlocal free scalar field $\hat \phi(x)$\footnote{In the case of field, we use $\hat ~$ notation to distinguish field operators from field configurations.} in a vacuum state $|0\rangle$ has the following correlation functions
\bea
\langle 0 | \hat \phi(x) |0\rangle&=&0\\
\langle 0 | \hat \phi(x)\hat \phi(y) |0\rangle&=&\int \frac{\dd^Dp}{(2\pi)^D} \tilde W(p) e^{i p\cdot (x-y)}\label{two_pointW},
\eea
where $D=d+1$ is the spacetime dimension and \footnote{$\theta(x)=1$ for $x>0$ otherwise $0$}
\beq
\tilde W(p)=2\pi \theta (p^0) \rho(-p^2).
\eeq
The function $\rho$, named the spectral function, has the following form \cite{Saravani:2015rva}
\beq\label{tilderho}
\rho(m^2)=\delta(m^2)+\tilde \rho(m^2)
\eeq
where $\tilde \rho(m^2)$ is a smooth function (free from singularities) and vanishes for $m^2<0$. \\
$\tilde W(p)$ is non-zero only when $p^0>0$ and $p^2\le 0$ ($p^2=p\cdot p$), i.e. it has only support inside and on the future light cone.
we can rewrite eq. \eqref{two_pointW} as  
\bea
\langle 0 | \hat \phi(x)\hat \phi(y) |0\rangle&&=\int \dd m^2  \rho(m^2) \int \frac{\dd^Dp}{(2\pi)^D} 2\pi \delta (p^2+m^2) \theta(p^0) e^{i p\cdot (x-y)}\notag\\
&&=\int \dd m^2 \rho(m^2) W_m(x,y) \label{two_point_rho},
\eea
where $W_m(x,y)$ is the two point correlation function of a (canonically normalized) local scalar field with mass $m$
\beq
W_m(x,y)= \int \frac{\dd^Dp}{(2\pi)^D} 2\pi \delta (p^2+m^2) \theta(p^0) e^{i p\cdot (x-y)}.
\eeq
The nonlocality scale is implicitly encoded in the function $\tilde \rho$. For practical purposes we can use $\tilde \rho (m^2)=l^2 e^{-m^2 l^2}$, where $l$ represents the nonlocality length parameter, tough our discussion in this manuscript does not assume this specific form. We only require that in the local limit, defined as $l\rightarrow 0$, $\tilde \rho(m^2)\rightarrow 0$. In this limit, the theory reduces to a local massless scalar field. In this sense, the scalar field $\hat \phi$ can be realized as the nonlocal modification of a massless scalar field.

Eqs. \eqref{two_point_rho} and \eqref{tilderho} play a crucial role throughout our discussion in this paper. They show that the two point correlation function of the scalar field $\hat \phi$ is the massless two point function plus a (continuum) sum over massive two point functions.
They also imply that $\hat \phi$ has massive and massless excitations. We can see this explicitly by the following field expansion which results into two point function \eqref{two_pointW}
\beq\label{field_expansion}
\hat \phi(x)=\int \frac{\dd^Dp}{(2\pi)^{D/2}}\sqrt{\tilde W(p)}\left(a_pe^{ip\cdot x}+a_p^\dagger e^{-i p\cdot x}\right)
\eeq
where all future directed non-spacelike momenta $p$ (those with $\tilde W(p)\neq 0$) contribute to the field expansion and $[a_p,a_q^\dagger]=\delta^{(D)}(p-q)$. Note that the creation and annihilation operators depend on both $p^0$ and $\bf p$ and the field expansion is an integral not limited to one mass shell. The vacuum state $|0\rangle$ is defined as
\beq
a_p|0\rangle=0~~\forall p,
\eeq
and the excited states are given by $a^\dagger$'s acting on the vacuum.

The Hamiltonian H and momentum $\bf P$ operators, defined as generators of time and spatial translations, are given as
\bea
&&\mbox{H}=\int \dd^Dp ~p^0 a_p^\dagger a_p,\\
&&{\bf P}=\int \dd^Dp ~{\bf p} ~a_p^\dagger a_p.
\eea
A state $|p\rangle\equiv a_p^\dagger |0\rangle$ is an eigenstate of H (with energy $p^0$) and $\bf P$ (with momentum $\bf p$), and as a result represents a 1-particle state with energy $p^0$ and momentum $\bf p$.

As we can see there are two types of excitations: Excitations $|p\rangle$ corresponding to the singularity of $\rho$ at $m=0$ and massive modes corresponding to the non-singular part of $\rho$ (namely $\tilde \rho$). We call the former ``isolated'' excitations and the latter ``continuum'' excitations \footnote{We believe this terminology better reflects the nature of these excitations. In previous papers on this subject, authors use a different terminology; for example, isolated and continuum modes are respectively called on-shell and off-shell modes in \cite{Saravani:2015rva}.}. By this terminology, a local field theory only possesses isolated excitations. The majority of our discussion in this manuscript concerns the continuum modes and their properties.

\subsection{Propagators}\label{propagator}

Before moving to the next section, let us introduce a number of quantities that are used throughout this paper. The two point function \eqref{two_point_rho} is a linear combination of local two point functions of massive fields, and as a result, it is straightforward to get
\bea
i\tilde \Delta(x,y)&&\equiv [\hat \phi(x),\hat \phi(y)]=\int dm^2 \rho(m^2)~i\Delta_m(x,y),\label{commutator}\\
i \tilde G_F(x,y)&&\equiv \langle 0 | \mbox{T}\hat \phi(x)\hat \phi(y) |0\rangle=\int \dd m^2  \rho(m^2)~iG_{F,m}(x,y),\label{time_ordered}
\eea
where 
\bea
i\Delta_m(x,y)&&=\int \frac{\dd^D p}{(2\pi)^D}2\pi \delta(p^2+m^2)\mbox{sgn}(p^0)e^{ip\cdot (x-y)},\\
iG_{F,m}(x,y)&&=\int  \frac{\dd^D p}{(2\pi)^D}\frac{-i}{p^2+m^2-i\epsilon}e^{ip\cdot (x-y)}
\eea
are, respectively, the commutator and the time-ordered two point function of a local scalar field with mass $m$ and $\epsilon$, as always, is a positive infinitesimal number taken to zero at the end of calculations. 

The retarded and advanced propagators can be obtained through $\tilde \Delta(x,y)=\tilde G_R(x,y)-\tilde G_A(x,y)$ as
\bea
\tilde G_R(x,y)&&=\tilde \Delta(x,y)\theta(x^0-y^0),\\
\tilde G_A(x,y)&&=-\tilde \Delta(x,y)\theta(y^0-x^0)=\tilde G_R(y,x).
\eea
This yields
\bea
\tilde G_R(x,y)&&= \int \dd m^2  \rho(m^2) G_{R,m}(x,y),\\
\tilde G_A(x,y)&&= \int \dd m^2  \rho(m^2) G_{A,m}(x,y),
\eea
where 
\bea
G_{R,m}(x,y)&&=\int  \frac{\dd^D p}{(2\pi)^D}\frac{-1}{p^2+m^2-i \epsilon p^0}e^{ip\cdot (x-y)},\\
G_{A,m}(x,y)&&=\int  \frac{\dd^D p}{(2\pi)^D}\frac{-1}{p^2+m^2+i \epsilon p^0}e^{ip\cdot (x-y)}
\eea
are, respectively, the retarded and advanced propagator of a local scalar field with mass $m$.
\begin{figure}
\includegraphics[width=\textwidth]{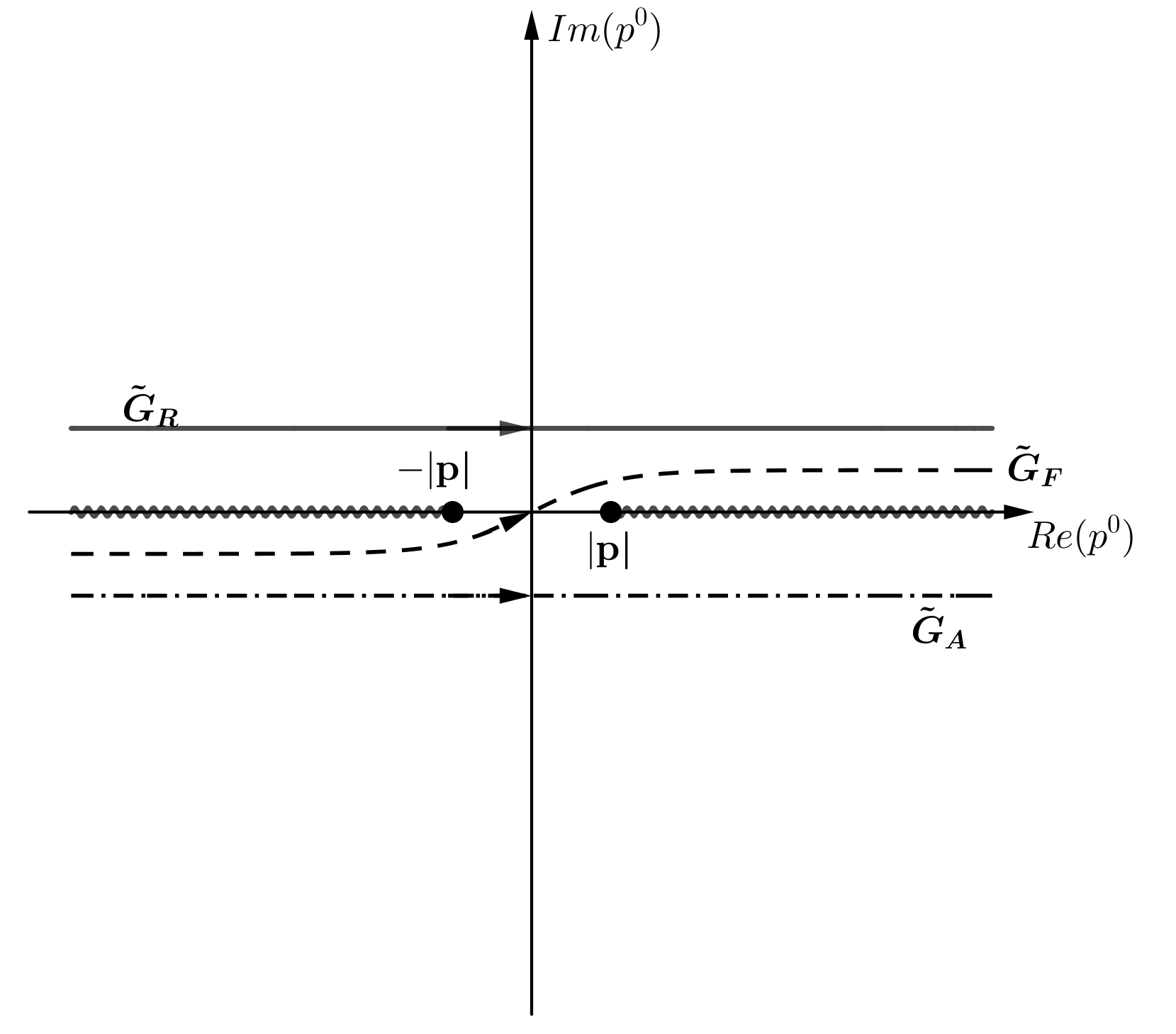}
\caption{Contours corresponding to the retarded (solid line), advanced (dot dashed) and Feynman (dashed) propagators in the complex plane of $p^0$. There is a branch cut on the real line corresponding to time-like momenta.}\label{contours}
\end{figure}

The eigenvalues of $\tilde G_R$, $\tilde G_A$ and $\tilde G_F$ in the momentum space have the following form
\beq\label{G_All}
\tilde G_\mathcal{X}(p)=\int \dd m^2 \rho(m^2)\frac{-1}{-(p^0+i \epsilon_\mathcal{X})^2+{\bf p}^2 +m^2}
\eeq
where $\mathcal{X}\in \{R,A,F\}$ and
\beq
\epsilon_R=\epsilon,\qquad \epsilon_A=-\epsilon,\qquad \epsilon_F=\epsilon p^0.
\eeq
For space-like momenta, the integrand in eq. \eqref{G_All} does not pass through a pole. Thus the result does not depend on the $i \epsilon$ prescription, and the values of all three propagators coincide. For time-like momenta, however, the integrand passes through a pole which results into a branch cut in the complex plane of $p^0$.
In other words, the three propagators correspond to different analytic continuations of the same function. In the complex plane of $p^0$, $\tilde G_R$ ($\tilde G_A$) corresponds to the values above (below) the branch cut and $\tilde G_F$ corresponds to the values below the branch cut for $p^0<0$ and above the branch cut for $p^0>0$ (see Figure \ref{contours}).

We can express quantities defined above in terms of the Fourier transform of the retarded propagator. In order to make a connection with the notation in \cite{Saravani:2015rva}, we define 
\beq
B(p)\equiv \left(\int \dd m^2  \rho(m^2) \frac{-1}{p^2+m^2-i \epsilon p^0}\right)^{-1}
\eeq
which only depends on $p^2$ and sgn$(p^0)$. Then, we can reexpress
\bea
&&\tilde G_R(x,y)=\int  \frac{\dd^D p}{(2\pi)^D}\frac{1}{B(p)}e^{ip\cdot (x-y)},\\
&&\tilde G_A(x,y)=\int  \frac{\dd^D p}{(2\pi)^D}\frac{1}{B^*(p)}e^{ip\cdot (x-y)},\\
&&\tilde G_F(x,y)=\int  \frac{\dd^D p}{(2\pi)^D}\left(\frac{\theta(p^0)}{B(p)}+\frac{\theta(-p^0)}{B^*(p)}\right)e^{ip\cdot (x-y)},\label{G_F_B}\\
&&i \tilde \Delta(x,y)=\int  \frac{\dd^D p}{(2\pi)^D}\left(\frac{1}{B(p)}-\frac{1}{B^*(p)}\right)e^{ip\cdot (x-y)},\\
&&\tilde W(p)=\frac{2\mbox{Im}B(p)\theta(p^0)}{|B(p)|^2}\label{BW}.
\eea
Operator $\tilde \Box$ defined as $\tilde \Box e^{ip\cdot x}=B(p)e^{ip\cdot x}$ acts as the starting point to define the nonlocal theory in \cite{Saravani:2015rva}. Note that unlike local field theories, $\tilde G_R(x,y)$, $\tilde G_A(x,y)$ and $\tilde G_F(x,y)$ are not Green's functions of a same operator. Each of them has a different inverse, a property that is tightly connected to the existence of continuum modes and the branch cut in Figure \ref{contours}.  
\section{Continuum modes}\label{continuum_modes}
The nonlocal theory described earlier contains massless and massive excitations at the same time. From eq. \eqref{two_point_rho} (as well as \eqref{commutator} and \eqref{time_ordered}), one may conclude simply that the theory is equivalent to having infinitely many independent local scalar fields for each value of mass. 
Let us first argue why this interpretation leads to inconsistent results.
For example, consider the Casimir force in the presence of this nonlocal field. The Casimir force corresponding to infinitely many local scalar fields, (infinitely) many of them with very small masses, would be simply infinite. Then, the mere finiteness of Casimir force in experiments would have ruled out this nonlocal theory. We explicitly calculate the Casimir force of the nonlocal field in section \ref{casimir} to probe the nonlocal modifications. 

We argue that, while the above interpretation has an element of truth to it, it is not the whole picture. The key difference between infinitely many local scalar fields and the nonlocal theory described here stems from their different spectral functions. The spectral functions of free local fields have only $\delta$-type singularities unlike the continuum modes of the nonlocal theory \eqref{tilderho}. We review a number of physical processes to address this point which further illuminate the physical interpretation of continuum modes and the role of $\tilde \rho(m^2)$ in its contribution to the spectral function.

\subsection{Particle production via source}\label{particle_production}
Consider the nonlocal scalar field in the vacuum coupled to a source $J(x)$, which is only present in a finite interval of time $[t_i,t_f]$. The quantum field picks up a c-number due to the source and evolves as (check Appendix \ref{source_response})
\beq\label{field_source1}
\hat \phi(x)=\int \frac{\dd^Dp}{(2\pi)^{D/2}}\sqrt{\tilde W(p)}\left(a^{in}_pe^{ip\cdot x}+a_p^{in\dagger} e^{-i p\cdot x}\right)+\int \dd^Dy~\tilde G_R(x,y)J(y).
\eeq 
At early times ($t<t_i$), the second term above vanishes and the field is assumed to be in its vacuum given by
\beq
a_p^{in}|0_{\mbox{in}}\rangle=0.
\eeq
At late times ($t>t_f$), the source is not present and
\beq
\int \dd^Dy~\tilde G_A(x,y)J(y)=0.
\eeq
Hence, the second term in \eqref{field_source1} can be expressed as
\bea
\int \dd^Dy~\tilde G_R(x,y)J(y)&=&\int \dd^Dy~\left(\tilde G_R(x,y)-\tilde G_A(x,y)\right)J(y)\notag\\
&=&\int \frac{\dd^D p}{(2\pi)^D}\bar J(p) \left(\frac{1}{B(p)}-\frac{1}{B^*(p)}\right)e^{ip\cdot x}\notag\\
&=&\int \frac{\dd^D p}{(2\pi)^D}\tilde W(p)\left(-i \bar J(p)e^{ip\cdot x}+i \bar J^*(p)e^{-ip\cdot x}\right),~~~~
\eea
where in the last line we have used eq. \eqref{BW} and
\beq
\bar J(p)\equiv \int \dd^D x~ J(x)e^{-i p\cdot x}.
\eeq
This results in the following expression for the field operator at late times
\bea
\hat \phi(x)&=&\int \frac{\dd^Dp}{(2\pi)^{D/2}}\sqrt{\tilde W(p)}\left[\left(a^{in}_p-i \frac{\sqrt{\tilde W(p)}\bar J(p)}{(2\pi)^{D/2}}\right)e^{ip\cdot x}+c.c\right]\notag\\
&=&\int \frac{\dd^Dp}{(2\pi)^{D/2}}\sqrt{\tilde W(p)}\left[a^{out}_pe^{ip\cdot x}+c.c\right],
\eea
where
\beq\label{a_in_out}
a_p^{out}\equiv a^{in}_p-i \frac{\sqrt{\tilde W(p)}\bar J(p)}{(2\pi)^{D/2}}.
\eeq
The scalar field at late time behaves as a free scalar field with a new vacuum state defined as
\beq
a_p^{out}|0_{\mbox{out}}\rangle=0.
\eeq
Starting from $|0_{\mbox{in}}\rangle$, the energy of the system at late times is given by
\bea
\mathcal{E}=\langle0_{\mbox{in}}|H^{out}|0_{\mbox{in}}\rangle&=&\langle0_{\mbox{in}}|\int \dd^D p~ p^0 a_p^{out \dagger}a_p^{out}|0_{\mbox{in}}\rangle\\
&=&\int \frac{\dd^D p}{(2\pi)^D} ~p^0\tilde W(p)|\bar J(p)|^2\label{energy_source_w}\\
&=&\int \dd m^2  \rho(m^2) \int\frac{\dd^d {\bf p}}{(2\pi)^d}\frac{1}{2} |\bar J(p)|^2\biggr\rvert_{p^0=E_{m, {\bf p}}}\\
&=&\int \dd m^2  \rho(m^2) \mathcal{E}_m,\label{energy_m}
\eea
where $E_{m, {\bf p}}=\sqrt{m^2 +{\bf p}^{2}}$ and  $\mathcal{E}_m$ is the energy of the system if the same source was coupled to a local massive scalar field of mass $m$
\beq\label{energy_local_m}
\mathcal{E}_m=\int\frac{\dd^d {\bf p}}{(2\pi)^d}\frac{1}{2} |\bar J(p)|^2\biggr\rvert_{p^0=E_{m,{\bf p}}}.
\eeq

Eq. \eqref{energy_m} shows that $\mathcal{E}$ is a linear combination of $\mathcal{E}_m$'s, each weighted by a factor of $\rho(m^2)$.
The nonlocal modification, which vanishes in the local limit and originates from the continuum modes is
\beq
\mathcal{E}-\mathcal{E}_{m=0}=\int \dd m^2 \tilde \rho(m^2) \mathcal{E}_m.
\eeq
While the nonlocal scalar field $\hat \phi$ contains infinitely many mass values, according to \eqref{energy_m}, the contribution from each mass is weighted by a factor of $ \rho(m^2)$ \footnote{To be precise, by a factor of $ \rho(m^2) \dd m^2$.}. This is very different from the naive result if each local field was coupled separately to the source; that would result in $\sum_m \mathcal{E}_m$.

\subsection{Decay rate and Unruh radiation}
The energy $\mathcal{E}$ in the particle production via a source term is an example of a quantity that depends linearly on the two point function of the scalar field. Notice that up to \eqref{energy_source_w} we have made no assumption about the field and $\tilde W$ represents the two point correlation function in the momentum space. Since $\mathcal{E}$ depends linearly on $\tilde W$, which is a linear combination of $W_m$'s, the final result \eqref{energy_m} is not surprising.

This linear dependence often occurs when quantities are calculated to the leading order in the nonlocal model. As an example, consider the spontaneous decay rate of an atom coupled to a nonlocal field, which has been studied in detail in \cite{Belenchia:2016sym}. The atom, modelled as a two level system in the excited state, is coupled to a nonlocal scalar field $\hat \phi$ in its vacuum. The atom decays by releasing a quanta of the scalar field. The crucial point is, to the leading order in couplings, the atom decay rate $\mathcal{F}$ depends linearly on the two point function of the nonlocal field. We represent it as
\beq
\mathcal{F}=O [\tilde W],
\eeq
where $O$ is a linear operation acting on $\tilde W$. Now if we define $\mathcal{F}_m$ to be the decay rate if the atom was coupled to a local massive scalar field $\hat \phi_m$ (exactly the same setup and interactions, just replacing $\hat \phi$ with $\hat \phi_m$), then
\beq
\mathcal{F}_m=O[W_m],
\eeq
then by linearity of $O$, we get
\beq\label{decay_rate_m}
\mathcal{F}=\int \dd m^2 \rho(m^2) \mathcal{F}_m.
\eeq
The nonlocal contribution, i.e. contribution from the continuum modes, is
\beq\label{decay_rate}
\mathcal{F}-\mathcal{F}_{m=0}=\int \dd m^2  \tilde \rho(m^2) \mathcal{F}_m.
\eeq
The decay rate is enhanced due to the existence of continuum modes, which provides more phase space for the atom to decay into, albeit by a small amount \footnote{In \cite{Belenchia:2016sym}, the authors show that the relative change is roughly given by $l^2 \Omega^2$ where $\Omega$ is the energy gap of the two level system and $l$ is the nonlocality length scale.}.  
Eq. \eqref{decay_rate} shows that the result is very different from an atom in contact with infinitely (many with small masses) local fields; that atom would instantly decay into its ground state.  

The same logic applies to the Unruh radiation. Consider a detector, a two level system prepared in its ground state, moving along a constant acceleration path, coupled to the nonlocal field $\hat \phi$ in its vacuum state. The detector absorption rate $\mathcal{A}$, to the leading order in couplings, depends linearly on the two point function $\tilde W$. Following the argument above, we conclude that
\beq\label{absorption_m}
\mathcal{A}=\int \dd m^2  \rho(m^2) \mathcal{A}_m,
\eeq
where $\mathcal{A}_m$ is the absorption rate of the same detector coupled to a massive local scalar field with mass $m$. 

Let us finish this section by summarizing the results:

\begin{itemize}

\item Continuum modes provide a bigger phase space for particle production, decay or absorption of energy. That is why in all cases above, we get an enhancement in the quantities calculated.

\item Unlike isolated modes, the contribution from a single mass of continuum modes is zero. Eqs. \eqref{energy_m}, \eqref{decay_rate_m} and \eqref{absorption_m} show that the isolated modes ($m=0$) provide a non-zero contribution (due to the $\delta$-function in the spectral function $\rho$). However, a zero measure set of masses of continuum modes makes no contribution.

\item Eqs. \eqref{energy_m}, \eqref{decay_rate_m} and \eqref{absorption_m} reduce to the massless scalar field results in the local limit.

\end{itemize}

\section{Path integral}\label{sec:path_integral}
In this section, we derive the path integral formulation of the nonlocal free theory. This further illustrates the role that $\tilde \rho(m^2)$ and the continuum modes play in the theory.

On one hand, it is rather straightforward to see what would be the right amplitude for the path integral of the free theory. We may choose the path integral to be a Gaussian integral resulting in the time-ordered two point correlation function \eqref{time_ordered}
\beq
\langle 0 | \mbox{T}\hat \phi(x)\hat \phi(y) |0\rangle=\frac{\int  \mathcal{D}\phi~\phi(x)\phi(y) e^{iS_{fr}[\phi]}}{\int  \mathcal{D}\phi~ e^{iS_{fr}[\phi]}}.\label{Feynman_two_point}
\eeq
We should choose the right quadratic action $S_{fr}[\phi]$ such that the above equality holds. Then, by the properties of Gaussian integrals and free quantum field theories, all higher correlation functions match. 

Let us find the correct form for the action $S_{fr}[\phi]$. For clarity we repeat the expression for the time-ordered two point function here
\bea
\langle 0 | \mbox{T}\hat \phi(x)\hat \phi(y) |0\rangle&=&i \tilde G_F(x,y),\\
\tilde G_F(x,y)&=& \int \dd m^2  \rho(m^2)\int  \frac{\dd^D p}{(2\pi)^D}\frac{-1}{p^2+m^2-i\epsilon}e^{ip\cdot (x-y)}\label{Feynman_mass}.
\eea
We define an operator $\tilde \Box_F$ to be the inverse of $\tilde G_F$, namely
\beq
\tilde \Box_F \tilde G_F(x,y)=\delta^{(D)}(x-y).
\eeq
The operator $\tilde \Box_F$ can be defined through its eigenvalues (see eq. \eqref{G_F_B})
\beq
\tilde \Box_F e^{ip\cdot x}=\left(\frac{\theta(p^0)}{B(p)}+\frac{\theta(-p^0)}{B^*(p)}\right)^{-1}e^{ip\cdot x}.
\eeq
Then,
\bea
S_{fr}[\phi]&\equiv &\int \dd^D x ~\frac{1}{2} \phi(x) (\tilde \Box_F \phi)(x)\label{Feynman_action}
\eea
provides the right amplitude for the path integral. 

In our discussion, we use $S_{fr}$ to define the path integral and one has to be careful with adding any extra physical interpretation to this quantity. For one thing, $\tilde \Box_F$ is not a real-valued operator and as a result, $S_{fr}$ is a complex number for a generic field configuration $\phi$. The imaginary part of $S_{fr}$ is always non-negative which makes the integral \eqref{Feynman_two_point} convergent.

In the next subsection, we derive the above result through a more canonical approach; by slicing the time interval into $N$ pieces and inserting the resolution of identity at each step. Before going into details, though, let us first explain why one may not expect to get a path integral in the form above.

Consider the vacuum to vacuum amplitude given by $\langle 0| e^{-i H t}|0 \rangle$. The traditional path integral derivation starts by separating $e^{-i H t}$ into $N$ pieces, inserting a resolution of identity at each step, calculating the matrix elements and taking the limit $N\rightarrow \infty$. For the local field theories, the time integral in the action comes from the $N$ time steps and spatial integrals come from the resolution of identity ($d$ dimensions) at each step, resulting in an action as a $D=d+1$ dimensional integral over spacetime configuration of the field.

However, this simple dimensional counting does not seem to work in the nonlocal theory. We still get the one dimensional time integral from $N$ time steps, but now the resolution of identity itself is a $D$ dimensional integral due to the presence of continuum modes (additional integral over mass). That is why one may expect to get a $D+1$ dimensional integral for the action at the end of the calculation, which clearly does not match \eqref{Feynman_action}. In what follows, we derive the action \eqref{Feynman_action} explicitly and show how one integral dimension is eliminated.

\subsection{Emergence of the dual picture}\label{dual_picture}
In order to derive the path integral, we discretize the mass integral in \eqref{two_point_rho} 
\beq
\langle 0 | \hat \phi(x)\hat \phi(y) |0\rangle=\sum_m \delta m^2  \rho(m^2) W_m(x,y),
\eeq
where $\delta m^2$ is the discretization step in $m^2$ and the equality holds when $\delta m^2\rightarrow 0$. By discretizing the mass, the derivation becomes clearer and easier to present. First, we express the field operator as
\bea
\hat \phi(x)&=&\sum_m \alpha_m\hat \phi_m(x),\\
\hat \phi_m(x)&=&\int \frac{\dd^d \bf{p}}{(2\pi)^{d/2}}\frac{1}{\sqrt{2E_{m,{\bf p}}}}\left(b_{m,{\bf p}}~e^{ip\cdot x}+c.c\right)\biggr\rvert_{p^0=E_{m,{\bf p}}},
\eea
where $\alpha_m=\sqrt{\rho(m^2)\delta m^2}$, $[b_{m_1,{\bf p}_1},b^\dagger_{m_2,{\bf p}_2}]=\delta_{m_1m_2}\delta^{(d)}({\bf p}_1-{\bf p}_2)$ and $\alpha_{m=0}=1$ due to the $\delta$-function on $m=0$. $\hat \phi_m$ operators are auxiliary fields defined here and we make use of their resemblance to local massive fields later in the calculation.

In the nonlocal theory, the eigenstates of the field operator $\hat \phi$ on a constant time slice are degenerate and no longer provide a complete basis. Let us explain this point further.  
First, note that $\hat \phi_{m_1}$ and $\hat \phi_{m_2}$ commute on a time slice: $[\hat \phi_{m_1}(t,{\bf x}),\hat \phi_{m_2}(t, {\bf y})]=0$. We define $|\phi_m\rangle$ as the eigenstate of $\hat \phi_m$ on a time slice $t_0$,
\beq
\hat \phi_m(t_0,{\bf x})|\phi_m\rangle=\phi_m({\bf x})|\phi_m\rangle.
\eeq
Then
\beq
|\Phi\rangle\equiv \tens{m}|\phi_m\rangle,\label{state_m}
\eeq
is an eigenstate of the field operator,
\beq\label{phi_eigenvalue}
\hat \phi(t_0,{\bf x})|\Phi\rangle=\sum_m \alpha_m \phi_m({\bf x})|\Phi\rangle.
\eeq
The above equation shows two states $|\Phi\rangle$ and $|\Phi'\rangle$ has the same eigenvalue for $\hat \phi$, as long as $\sum_m \alpha_m \phi_m({\bf x})=\sum_m \alpha_m \phi'_m({\bf x})$, i.e. $\hat \phi$ is degenerate. The complete basis for the Hilbert space is given by the eigenstates of $b_{m,{\bf p}}$ operators or equivalently, the eigenstates of $\hat \phi_m$ operators on a constant time slice, given by eq. \eqref{state_m}.   
Then, the resolution of identity is 
\beq\label{identity}
I=\int \mathcal{D}\Phi |\Phi\rangle\langle\Phi|,
\eeq
where $\mathcal{D}\Phi=\prod_m \mathcal{D}\phi_m$ and $\phi_m$'s are field configurations on a time slice.

The vacuum state defined as $b_{m,{\bf p}}|0\rangle=0~~ \forall ~m, {\bf p}$ can be expressed as the tensor product of vacuums associated to each mass
\beq\label{vacuum_m}
|0\rangle=\tens{m} |0_m\rangle,
\eeq
where $|0_m\rangle$ is defined as
\beq
b_{m,{\bf p}}|0_m\rangle=0~~ \forall  {\bf p}.
\eeq
Finally, the Hamiltonian of the theory in terms of $b$ operators is
\beq\label{Hamiltonian_m}
H=\sum_m \int \dd^d {\bf p}~ E_{m,{\bf p}}~b^\dagger_{m,{\bf p}}~ b_{m,{\bf p}}=\sum_m H_m,
\eeq
where 
\beq
H_m=\int \dd^d {\bf x}~:\frac{1}{2}(\partial_t\hat \phi_m)^2+\frac{1}{2}({\bf \nabla}\hat\phi_m)^2+\frac{1}{2}m^2\hat \phi_m^2:
\eeq
is the (normal ordered) free Hamiltonian associated with the massive scalar field $\hat \phi_m$.

Let us start with the vacuum-to-vacuum amplitude $\langle 0 | e^{-i H t}|0\rangle$. Using eqs. \eqref{vacuum_m} and \eqref{Hamiltonian_m}, we get
\beq
\langle 0 | e^{-i H t}|0\rangle=\prod_m \langle 0_m | e^{-i H_m t}|0_m\rangle.
\eeq
$\langle0_m | e^{-i H_m t}|0_m\rangle$ is the vacuum-to-vacuum amplitude of a local massive scalar field with mass $m$. The path integral presentation of this term is
\beq\label{vv_amplitude}
\langle 0 | e^{-i H t}|0\rangle=\prod_m \langle 0_m | e^{-i H_m t}|0_m\rangle=\prod_m\int \mathcal{D}\phi_m~e^{iS_m[\phi_m]}=\int \prod_m \mathcal{D}\phi_m~e^{i\sum_m S_m},
\eeq
where $\phi_m$'s are spacetime field configurations and 
\beq
S_m=\int \dd^Dx ~\frac{1}{2}\phi_m(\Box-m^2)\phi_m.
\eeq
The right hand side of \eqref{vv_amplitude} is a product of vacuum-to-vacuum amplitudes of local massive fields and the contributions from different masses are completely disentangled. This is essentially what we argued in the last paragraph before Sec. \ref{dual_picture}. This result is very different from \eqref{Feynman_action}: there is an additional sum over the mass parameter in the action and the path integral includes integration over all $\phi_m$ configurations rather than a single field configuration.

In order to realize how to deal with the additional sum over mass, let us consider the time ordered two point function $\langle 0|T\hat\phi(x)\hat \phi(y)|0\rangle$. Following the above steps and using \eqref{state_m} and \eqref{phi_eigenvalue}, we get
\beq\label{path_integral_m}
\langle 0|T\hat\phi(x)\hat \phi(y)|0\rangle=\frac{\int \prod_m \mathcal{D}\phi_m ~~  \left[\sum_m \alpha_m \phi_m(x)\right]\left[\sum_m \alpha_m \phi_m(y)\right]e^{i \sum_m S_m}}{\int \prod_m \mathcal{D}\phi_m ~~ e^{i \sum_m S_m}}.
\eeq
Again, the above result is very different from \eqref{Feynman_two_point}; it is a path integral over many scalar field configuration instead of one. 

However, eq. \eqref{path_integral_m} can be simplified more. Notice that only the combination $\sum_m \alpha_m \phi_m$ appears in the integrand of the path integral. We can perform the integrals in \eqref{path_integral_m}, while keeping the combination $\sum_m \alpha_m \phi_m$ fixed. This can be done as follows:
\bea
&&\int \prod_m \mathcal{D}\phi_m ~~ \left(\sum_m \alpha_m \phi_m(x)\right)\left( \sum_m \alpha_m \phi_m(y)\right)e^{i \sum_m S_m} \label{Feynman_m_derivation1}\\
&&=\int \mathcal{D}\phi \int \prod_m \mathcal{D}\phi_m ~~ \phi(x)\phi(y) ~\delta\left(\phi-\sum_m \alpha_m \phi_m\right)e^{i \sum_m S_m} \\
&&=\mathcal{N}\int  \mathcal{D}\phi~\phi(x)\phi(y) e^{iS_\delta [\phi]}\label{Feynman_m_derivation2}.
\eea
$\mathcal{N}$ is a numerical constant which will be cancelled by the same contribution from the denominator in \eqref{path_integral_m}. $S_\delta$ is (check Appendix \ref{path_integral_proof} for proof)
\beq
S_\delta[\phi]=\int \dd^Dx ~\frac{1}{2}\phi(x)(\Box_\delta \phi)(x),
\eeq
where 
\beq
\Box_\delta^{-1}=\sum_m \frac{\alpha_m^2}{\Box-m^2+i\epsilon}=\sum_m \frac{ \rho(m^2)\delta m^2}{\Box-m^2+i\epsilon}.
\eeq
The above expression becomes the Feynman propagator \eqref{Feynman_mass} in the limit $\delta m^2\rightarrow 0$. As a result, $\Box_\delta$ reduces to $\tilde \Box_F$ in this limit and we recover the path integral \eqref{Feynman_two_point}.

The key property that allows us to extend this result to all higher order correlation functions is the fact that the integrand of the path integral only involves $\sum_m \alpha_m \phi_m$. As illustrated above, we can always integrate out $\phi_m$ fields, while keeping this combination fixed. 

\section{Dual picture of interactions}

The Feynman path integral provides a natural way to introduce interactions to the theory by adding an interaction term to the free action \eqref{Feynman_action}
\beq\label{interaction_nonlocal}
S_{int}[\phi]=-\int \dd^D x~~  V(\phi).
\eeq
In principle, one can compute the correlation functions or derive the Feynman rules to compute scattering amplitudes using the path integral of the nonlocal interacting theory. On the other hand, the dual description of the path integral \eqref{path_integral_m} in terms of $\phi_m$'s allows one to work with local actions. In oder to include the interaction term, we only need to add the following expression
\beq\label{interaction_alpha_m}
S_{int}\left[\sum_m\alpha_m \phi_m\right]=-\int \dd^D x ~ V\left(\sum_m\alpha_m \phi_m\right)
\eeq 
to the sum of the local field actions in \eqref{path_integral_m}. Then the above contribution becomes the interaction term \eqref{interaction_nonlocal}, after integrating out $\phi_m$'s, as follows
\bea
\int \prod_m \mathcal{D}\phi_m && \delta\left(\phi-\sum_m \alpha_m \phi_m\right)exp\left(i \sum_m S_m[\phi_m]+i S_{int}\left[\sum_m\alpha_m \phi_m\right]\right)\notag\\
&&=e^{i S_{int}[\phi]}\int \prod_m \mathcal{D}\phi_m ~~ ~\delta\left(\phi-\sum_m \alpha_m \phi_m\right)e^{i\sum_mS_m}\notag\\
&&=\mathcal{N}e^{i\left(S_{fr}[\phi]+S_{int}[\phi]\right)}.
\eea
We discuss the importance of \eqref{interaction_alpha_m} in the following section.

\subsection*{Interpretation of the dual picture}
Let us take a look again at the particle production via a source term (Section \ref{particle_production}). In the dual picture, the nonlocal field $\phi$ is replaced by the local fields $\phi_m$ with the specific combination $\sum_m \alpha_m \phi_m$ in the interaction terms. This gives a physical intuition on why we arrived at the result \eqref{energy_m}.

The source term can be introduced to the action via 
\beq
S_J=-\int \dd^Dx ~ J(x)\phi(x).
\eeq
Using the result of the previous section, we know this term translates into the following in terms of $\phi_m$ fields
\beq
S_J=-\int \dd^D x \sum_m \alpha_m \phi_m(x) J(x).
\eeq
This means that each local scalar field $\phi_m$ is coupled to a source via 
\beq
S_J^{(m)}=-\int \dd^Dx~ \alpha_m J(x)\phi_m(x).
\eeq
In the dual picture, the source $J$ is coupled to the scalar fields $\phi_m$ with an additional coupling constant $\alpha_m$. This means that the energy produced by the source in the field $\phi_m$ is $\alpha_m^2 \mathcal{E}_m$ (as defined in eq. \eqref{energy_local_m}). The total energy is 
\beq
\mathcal{E}=\sum_m \alpha_m^2 \mathcal{E}_m=\sum_m \rho(m^2) \delta m^2 \mathcal{E}_m,
\eeq 
which coincides with \eqref{energy_m} when $\delta m^2\rightarrow 0$.

Here we see the important role of $\alpha_m$ factors: in terms of the local fields $\phi_m$, all the interaction terms get an additional coupling constant $\alpha_m$ for the scalar field $\phi_m$. Note that $\alpha_m$ is an infinitesimal number except for $m=0$. That is why the massless field, on its own, has a non-zero contribution to the total energy ($\alpha_{m=0}=1$) in the example above, while each massive field contributes infinitesimally to the total energy. Only a collection of massive fields (a set with non-zero measure in mass) contributes to the total energy. 

At this point, it should be clear where the qualitative difference between isolated and continuum modes originates from. 
In the dual picture, the local fields $\phi_m$ are always accompanied by $\alpha_m$ in the interaction terms. For massive fields (continuum modes), $\alpha_m$ is an infinitesimal number. As a result, contributions from massive fields are suppressed by this factor. That is why only a {\it continuum} of continuum modes can counterbalance this suppression factor and contribute to physical observables. Unlike continuum modes, there is no suppression factor for isolated modes.

\section{S-matrix amplitudes}\label{S-matrix}
\begin{figure}[t!]
	\begin{subfigure}{0.45\textwidth}
	\includegraphics[width=\textwidth]{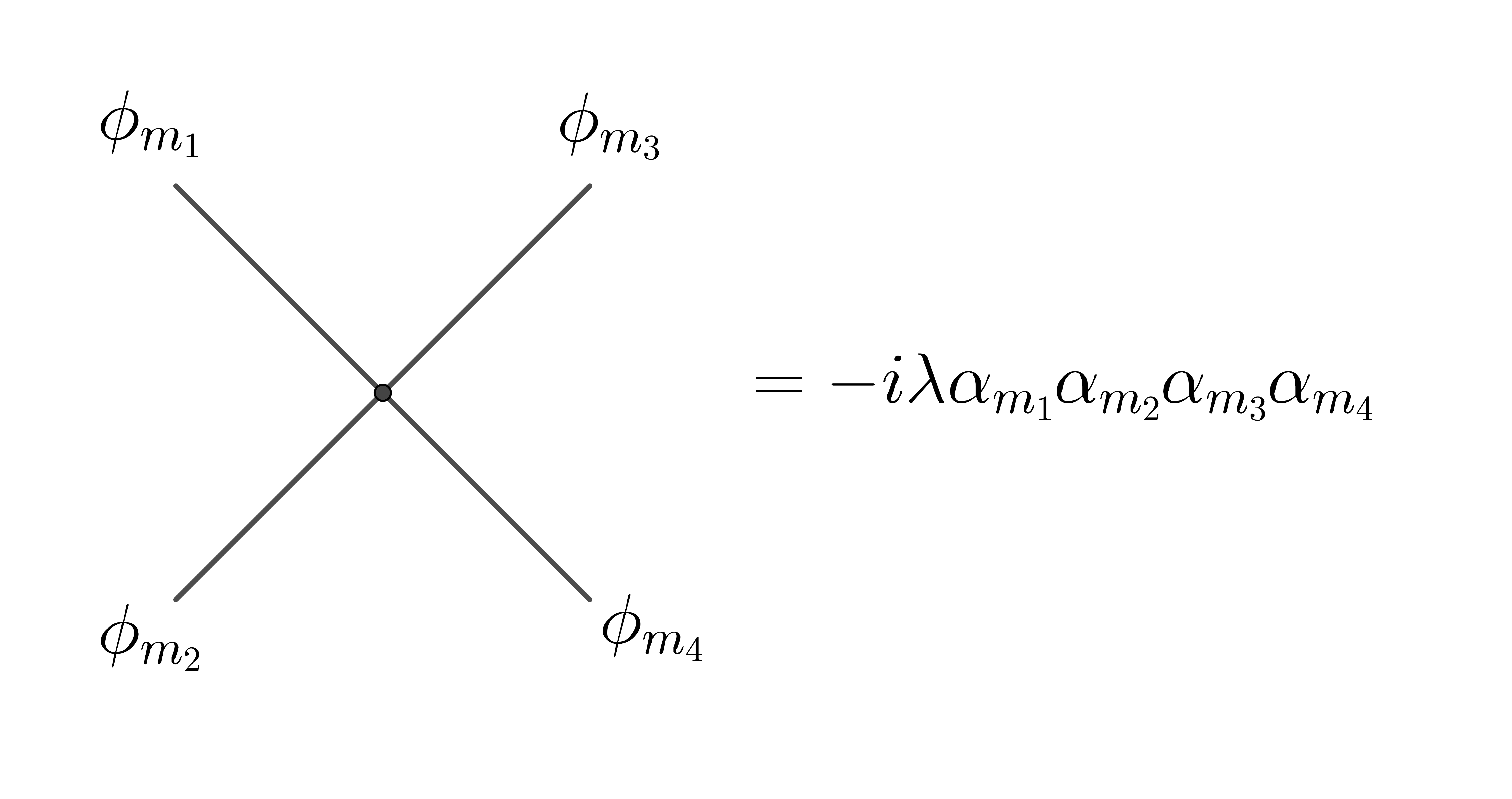}
	\caption{The interaction vertex amplitude of $\frac{\lambda}{4!}\phi^4$ before any simplification. For each $\phi_m$ field, an extra factor of $\alpha_m$ is included.}
	\label{fig:vertex}
	\end{subfigure}\hfill
	\begin{subfigure}{0.45\textwidth}
	\includegraphics[width=\textwidth]{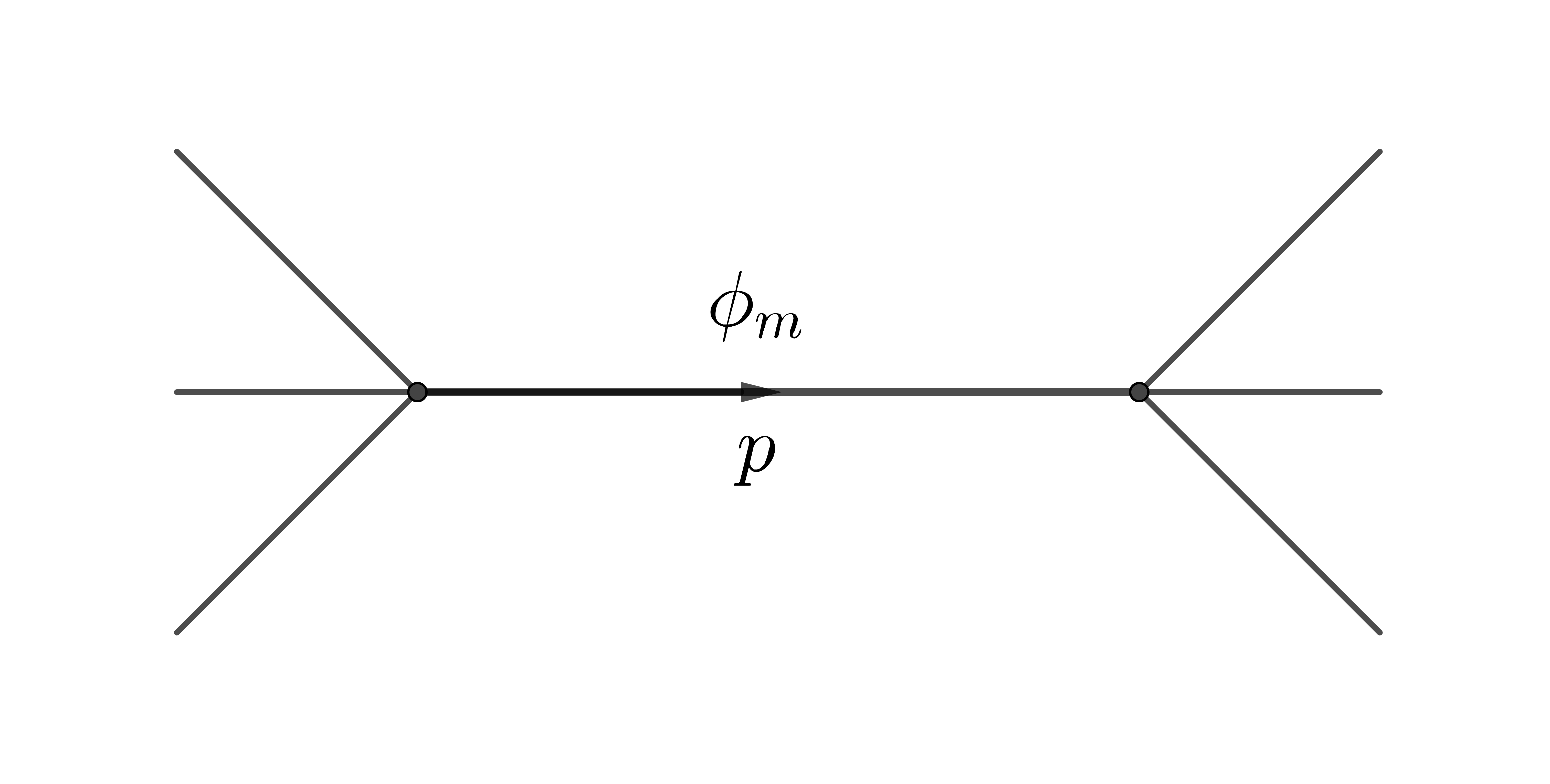}
	\caption{An internal line connecting two vertices. There are exactly similar Feynman diagrams to the above, just replacing $\phi_m$ with $\phi_{m'}$.}
	\label{fig:internal_line}
	\end{subfigure}
	\caption{}
\label{fig:S-matrix}
\end{figure}

It is rather straightforward to derive the S-matrix amplitudes of a nonlocal interacting theory with the help of the dual picture. Remember that in the dual picture, the interaction term $V(\phi)$ is replaced by $V(\sum_m \alpha_m \phi_m)$. In principle, there are infinitely many Feynman diagrams, even for a simple scattering process with one internal line, to be summed over. However, we claim that the Feynman rules simplifies to the following two steps:
\begin{enumerate}
\item Consider a local field with the same interaction and draw all Feynman diagrams for this theory.
\item There are only two modifications to evaluating each diagram: a) propagators associated to each internal line is $\tilde G_F(p)$ and b) the contribution from an external line with mass $m$ gets an additional factor $\alpha_m$. 
\end{enumerate}
Let us argue why.
Since the interaction term in the dual picture is $V(\sum_m \alpha_m \phi_m)$, each vertex in the Feynman diagram gets additional $\alpha_m$ factors, depending on the fields attached to the vertex (e.g. see Fig. \ref{fig:vertex} for $V(\phi)=\frac{\lambda}{4!}\phi^4$). 
Moreover, for each internal $\phi_m$ line connecting two vertices (Fig. \ref{fig:internal_line}), there are exactly similar Feynman diagrams with $\phi_{m'}$ as the internal line. The difference between the values of these diagrams are a) the internal line propagator $\frac{-i}{p^2+m^2-i\epsilon}$ and b) the two $\alpha_m$ factors from the two vertices connected to the internal line. Summing over all these diagrams tells us that the propagator associated to the internal line is modified to
\beq
\sum_m \alpha_m^2\frac{-i}{p^2+m^2-i\epsilon}=-i \tilde G_F(p).
\eeq
After performing the above for every internal line in the Feynman diagram, we have absorbed all $\alpha_m$ factors of the internal lines. The remaining $\alpha_m$ factors are associated to the external lines.

\section{The Casimir effect}\label{casimir}
We end our discussion on the nonlocal field theory by a phenomenological question: what is the modification to the Casimir force for the scalar field $\phi$ with the nonlocal dynamic as described in this paper?

Let us start by establishing notation. A spacetime point $x=(x^0,X,{\bf x}_T)$ where $X$ represents the direction where boundary conditions are imposed and ${\bf x}_T$ is the $d-1$ dimensional transverse space. The field is required to vanish at $X=0$ and $X=a$
\beq\label{casimir_boundary_condition}
\phi(x^0,X=0,{\bf x}_T)=\phi(x^0,X=a,{\bf x}_T)=0.
\eeq
Before going into the details of the calculation, let us use the intuition we have gained so far about the nonlocal field $\phi$ to anticipate whether the Casimir force is strengthened or weakened by the presence of nonlocality. 

From the discussion in the previous sections, we learned that, in the dual picture, $\phi$ is replaced by a collection of local scalar fields $\phi_0+\sum_{m\neq 0}\alpha_m\phi_m$, where we have separated massless from massive fields in this expression. The boundary condition \eqref{casimir_boundary_condition} requires this combination (and not individual fields) to vanish on $X=0$ and $X=a$
\beq\label{boundary_condition_nonlocal}
\left(\phi_0+\sum_{m\neq 0}\alpha_m\phi_m\right)\biggr\rvert_{X=0}=\left(\phi_0+\sum_{m\neq 0}\alpha_m\phi_m\right)\biggr\rvert_{X=a}=0.
\eeq
Note that in the local limit $\alpha_m= \delta_{m,0}$, the boundary condition is only imposed on the massless scalar field
\beq\label{boundary_condition_local}
\phi_0\biggr\rvert_{X=0}=\phi_0\biggr\rvert_{X=a}=0.
\eeq
The boundary condition \eqref{boundary_condition_nonlocal} implies that the massless scalar field $\phi_0$ can fluctuate and be non-zero on the boundaries as long as its fluctuations are cancelled out by the combination of massive fields. In other words, the massless scalar field has a ``weaker'' boundary condition compared to \eqref{boundary_condition_local} in the local limit. As a result, we expect the Casimir force in the nonlocal case to be smaller than that of a massless scalar field.

Now let us move on to explicitly calculate the Casimir force in the nonlocal theory. We follow the method discussed in Ref. \cite{Ambjorn:1981xw}. In this method, we calculate the Casimir energy by deriving the partition function of the theory and reading off the zero point energy from the partition function. In order to do so, we start with the path integral of the free theory (no boundary and no interaction) and perform a Wick rotation to get the partition function
\beq
Z_0=\int \mathcal{D}\phi~ e^{-S_0},
\eeq
where 
\beq
S_0=\int \frac{1}{2} \dd^D x ~\phi  \tilde \Box_E \phi
\eeq
and $\tilde \Box_E$ is the Wick rotation of $\tilde \Box_F$. In order to define $\tilde \Box_E$, consider the Feynman propgator
\beq
\tilde G_F(p)=\int \dd m^2 \rho(m^2) \frac{1}{-(p^0)^2+P^2+{\bf p}_T^{~2} +m^2-i\epsilon},
\eeq
where $p=(p^0,P,{\bf p}_T)$. After Wick rotation ($t\rightarrow i t$ and $p^0\rightarrow i p^0$), this yields
\beq\label{GFE}
\tilde G_E(p)=\int \dd m^2  \rho(m^2) \frac{1}{(p^0)^2+P^2+{\bf p}_T^{~2} +m^2}
\eeq
and $\tilde \Box_E$ is defined as the inverse of the propagator
\beq
\tilde \Box_E e^{ip\cdot x}=\frac{1}{\tilde G_E(p)}e^{ip\cdot x},
\eeq
where $p\cdot x$ is now the Euclidean inner product. 

After imposing the boundary conditions, translation invariance is only broken in the $X$ direction. That is why we make use of the translation invariance in the remaining $d$ directions and perform the (inverse) Fourier transform of \eqref{GFE} only in the $X$ direction to get
\beq\label{G_0}
G(\kappa,X-Y)\equiv \int \frac{\dd P}{2\pi} \tilde G_E(p) e^{iP(X-Y)}= \int dm^2  \rho(m^2) \frac{1}{2\sqrt{\kappa^2+m^2}}e^{-\sqrt{\kappa^2+m^2}|X-Y|},
\eeq
where $\kappa=\sqrt{(p^0)^2+{\bf p}_T^{~2}}$.

In order to introduce the boundary conditions \eqref{casimir_boundary_condition}, consider the following theory
\beq\label{Z_interacting}
Z=\int \mathcal{D}\phi~e^{-S},
\eeq
where 
\bea
S&&=\int \dd^D x ~\frac{1}{2} \phi \tilde \Box_E \phi+\frac{1}{2}\lambda_1 \delta(X) \phi(x)^2+\frac{1}{2}\lambda_2 \delta(X-a) \phi(x)^2,\notag\\
&&=\int \dd^D x \frac{1}{2} \phi\left(\tilde \Box_E+V_1+V_2\right)\phi\label{Euclid_action_interacting}
\eea
and we have defined 
\bea
(V_1\phi)(x)&\equiv& \int \dd^D y \lambda_1 \delta(X)\delta^{(D)}(x-y)\phi(y),\\
(V_2\phi)(x)&\equiv& \int \dd^D y \lambda_2 \delta(X-a)\delta^{(D)}(x-y)\phi(y).
\eea
We use the above notation to treat $V_1(x,y)=\lambda_1\delta(X)\delta^{(D)}(x-y)$ and $V_2(x,y)=\lambda_2\delta(X-a)\delta^{(D)}(x-y)$ as operators acting on the field. 

In the limit $\lambda_1,\lambda_2\rightarrow \infty$, only the field configurations where $\phi(x)|_{X=0}=\phi(x)|_{X=a}=0$ contribute to the partition function. As a result, we are effectively imposing the boundary conditions \eqref{casimir_boundary_condition} by introducing $V_1$ and $V_2$ potentials and the Casimir energy can be interpreted as the energy change in the system due to the potentials in the limit $\lambda_1,\lambda_2\rightarrow \infty$.

Since the Lagrangian \eqref{Euclid_action_interacting} is translation invariant in ``time'' and transverse spatial directions, we can perform Fourier transform in these $d$ directions and focus only on the non-trivial $X$ direction. After the Fourier transform
\bea
V_1(X,Y)&=&\lambda_1 \delta(X)\delta(X-Y),\\
V_2(X,Y)&=&\lambda_2\delta(X-a)\delta(X-Y).
\eea

Evaluating the integral \eqref{Z_interacting}, the partition function is given by
\beq\label{ln_z}
\ln Z=-\frac{1}{2}\ln \mbox{Det} \left(\tilde \Box_E+V_1+V_2\right)=-\frac{1}{2}\mbox{Tr}\ln(\tilde \Box_E+V_1+V_2).
\eeq

In Appendix \ref{app_casimir}, we show that the zero point energy per unit area of the transverse directions, $\mathcal{E}$, is given by
\beq\label{casimir_final}
\mathcal{E}=\frac{1}{2}\frac{S_{d-1}}{(2\pi)^d}\int_0^{\infty}\dd\kappa ~\kappa^{d-1} \ln\left[1-\left(\frac{G(\kappa,a)}{G(\kappa,0)}\right)^2\right],
\eeq
where $S_{d-1}=\frac{2\pi^{d/2}}{\Gamma(d/2)}$ is the area of a $d$-dimensional unit sphere and we have taken the limit $\lambda_1,\lambda_2\rightarrow \infty$.
Eq. \eqref{casimir_final} allows us to calculate the Casimir energy for a particular choice of $\tilde \rho(m^2)$. A theoretically motivated choice is $\tilde \rho(m^2)=l^2 e^{-m^2 l^2}$ (see \cite{Saravani:2015rva,Belenchia:2016sym}), where $l$ represents the nonlocality length scale. After substituting this value in eqs. \eqref{G_0} and \eqref{casimir_final}, the leading order correction to the Casimir energy reduces to a rather simple expression  
\beq\label{casimir_expression}
\mathcal{E}=\mathcal{E}_0\left(1-d\sqrt{\pi}\frac{l}{a}\right)+\mathcal{O}(l^2/a^2),
\eeq
where $\mathcal{E}_0$ is the Casimir energy of a massless scalar field with absorbing plates separated by distance $a$  
\beq
\mathcal{E}_0=-\frac{ \zeta (D) \Gamma \left(\frac{D}{2}\right)}{2^D\pi^{D/2}a^d}.
\eeq
Eq. \eqref{casimir_expression} shows that the nonlocal contribution appears with an opposite sign, as we argued at the beginning of this Section.

\section{Concluding remarks}
A class of nonlocal field theories, as we dubbed NAQFT, possesses a continuum of massive excitations. As one of the most important features of this class of theories, we studied the role of the continuum modes in detail. The path integral formulation of this theory was derived and led to a dual picture in terms of local fields. The dual picture highlights how the continuum modes of the nonlocal field behave. We have also derived the Feynman rules for evaluating the S-matrix amplitudes using the dual picture.

Let us end this manuscript with two important points for future studies:
\begin{enumerate}
\item Our discussion in this manuscript was entirely limited to scalar field theories. However, the dual picture provides a path to extension to fermions. The key idea is formalized in eq. \eqref{phi_integrate_out}; the nonlocal theory can be understood as integrating out a collection of local massive and massless fields while only keeping a linear combination of them. There is no obstruction to generalize this idea to fermions. However, extension to gauge fields is not as straightforward, since a mass term breaks gauge invariance. 
\item The dual picture allows us to address the renormalizability of these theories. In the nonlocal picture, the principles of how to renormalize the theory are not clear. However, the dual picture is a local field theory and we have a thorough understanding of how renormalization works.  
\end{enumerate}

\subsection*{Acknowledgements}
We would like to thank Ted Jacobson and Rafael Sorkin for numerous discussions. We are grateful to Robert Benkel, Nicola Franchini and Sumati Surya for a critical reading of the earlier version of this draft. M.S. is supported by the Royal Commission for the Exhibition of 1851.

\bibliographystyle{jhep}
\bibliography{NLQFT}


\appendix 
\section{Response of the nonlocal field to a source}\label{source_response}

Here, we calculate the response of the nonlocal field to a source term and show it corresponds to eq. \eqref{field_source1}. We add the source term via the path integral introduced in the Section \ref{sec:path_integral}
\beq\label{path_integral_source}
S=\int \dd^Dx ~\frac{1}{2}\phi \tilde \Box_F\phi-J\phi
\eeq
and show that eq. \eqref{field_source1} produces the same correlation functions as the path integral. We get the following correlation functions by the path integral:
\bea
\langle \phi(x)\rangle&&=(\tilde G_FJ)(x),\label{one_point_path_integral}\\
\langle \phi(x)\phi(y)\rangle&&=i\tilde G_F(x,y)+(\tilde G_FJ)(x)(\tilde G_FJ)(y).\label{two_point_path_integral}
\eea
Higher point correlation functions can be derived using the properties of Gaussian integrals. 

Now, we show that the field expansion \eqref{field_source1} produces the same correlation functions. Note that the Feynman path integral yields in-out time-ordered correlation functions. We start by the field expansion \eqref{field_source1} to calculate the in-out field operator value
\bea
\frac{\langle 0_{\mbox{out}}|\hat \phi(x)|0_{\mbox{in}}\rangle}{\langle 0_{\mbox{out}}|0_{\mbox{in}}\rangle}&&= (\tilde G_RJ)(x)+\int \frac{\dd^Dp}{(2\pi)^{D/2}}\sqrt{\tilde W(p)}\frac{\langle 0_{\mbox{out}}|a_p^{in\dagger}|0_{\mbox{in}}\rangle}{\langle 0_{\mbox{out}}|0_{\mbox{in}}\rangle}   e^{-i p\cdot x}\notag\\
&&=(\tilde G_RJ)(x)-i\int \frac{\dd^Dp}{(2\pi)^{D}}\tilde W(p)\bar J^*(p)   e^{-i p\cdot x}\notag\\
&&=(\tilde G_RJ)(x)-i\int \frac{\dd^Dp}{(2\pi)^{D}}\tilde W(-p)\bar J(p)   e^{i p\cdot x}\notag\\
&&=(\tilde G_FJ)(x),
\eea
where in the second line we used eq. \eqref{a_in_out}. This matches \eqref{one_point_path_integral}. 

Rather straightforward calculations yield
\beq
\frac{\langle 0_{\mbox{out}}|\hat \phi(x)\hat \phi(y)|0_{\mbox{in}}\rangle}{\langle 0_{\mbox{out}}|0_{\mbox{in}}\rangle}=(\tilde G_FJ)(x)(\tilde G_FJ)(y)+\int \frac{\dd^D p}{(2\pi)^D}\tilde W(p)e^{ip\cdot(x-y)}.
\eeq
After time ordering, we get
\beq
\frac{\langle 0_{\mbox{out}}|T\hat \phi(x)\hat \phi(y)|0_{\mbox{in}}\rangle}{\langle 0_{\mbox{out}}|0_{\mbox{in}}\rangle}=(\tilde G_FJ)(x)(\tilde G_FJ)(y)+i\tilde G_F(x,y).
\eeq
This matches \eqref{two_point_path_integral}. Using the properties of the free theory (field commutator is still a c-number), all higher order correlation functions of the field expansion \eqref{field_source1} and the path integral \eqref{path_integral_source} are equal as well. 

\section{Path integral proof}\label{path_integral_proof}
Here, we prove 
\beq\label{phi_integrate_out}
\int \prod_m \mathcal{D}\phi_m ~ ~\delta\left(\phi-\sum_m \alpha_m \phi_m\right)e^{i \sum_m S_m} \\
= \mathcal{N}e^{iS_\delta [\phi]},
\eeq
where 
\beq
S_\delta[\phi]=\int \dd^Dx ~\frac{1}{2}\phi\Box_\delta \phi
\eeq
and
\beq
\Box_\delta^{-1}=\sum_{m} \frac{\alpha_m^2}{\Box-m^2+i\epsilon}.
\eeq
In the rest of this section, we drop the integration constant $\mathcal{N}$. First, we evaluate the following expression
\beq
I[\phi]=\int \mathcal{D}\psi\mathcal{D}\chi ~~\delta\left(\phi-\psi- \alpha \chi\right)exp\left(i \int \frac{1}{2}\psi O_\psi \psi+\frac{1}{2}\chi O_\chi\chi\right),
\eeq
where operators $O_\psi$ and $O_\chi$ are functions of the d'Alembertian. Performing the integrals, first on $\psi$ and then on $\chi$ yield
\beq\label{master_term}
I[\phi]=e^{i\int \frac{1}{2}\phi O_\phi\phi},
\eeq
where 
\beq
O_\phi^{-1}=O_\psi^{-1}+\alpha^2O_\chi^{-1}.
\eeq
Now we prove \eqref{phi_integrate_out} by induction on the number of $\phi_m$ fields ($n$). Eq. \eqref{master_term} proves \eqref{phi_integrate_out} for $n=2$ when $O_\psi=\Box+i\epsilon$ and $O_\chi=\Box-m^2+i\epsilon$. Assuming \eqref{phi_integrate_out} is correct for $n$, we prove it for $n+1$. Consider the left hand side of \eqref{phi_integrate_out} for $n+1$ fields
\bea
&&I_{n+1}=\int \prod_{k=0}^n \mathcal{D}\phi_{m_k} ~ ~\delta\left(\phi-\sum_{k=0}^n \alpha_{m_k} \phi_{m_k}\right)e^{i \sum_{k=0}^n S_{m_k}}\notag\\
&&=\int\mathcal{D}\phi_{m_n}e^{iS_{m_n}}\int \prod_{k=0}^{n-1} \mathcal{D}\phi_{m_k} ~ ~\delta\left(\phi-\alpha_{m_n} \phi_{m_n}-\sum_{k=0}^{n-1} \alpha_{m_k} \phi_{m_k}\right)e^{i \sum_{k=0}^{n-1} S_{m_k}}\notag\\
&&=\int\mathcal{D}\phi_{m_n}\mathcal{D}\psi\delta(\phi-\psi-\alpha_{m_n} \phi_{m_n})e^{iS_{m_n}}\int \prod_{k=0}^{n-1} \mathcal{D}\phi_{m_k} ~ ~\delta\left(\psi-\sum_{k=0}^{n-1} \alpha_{m_k} \phi_{m_k}\right)e^{i \sum_{k=0}^{n-1} S_{m_k}}.\notag\\
\eea  
Using the assumption of induction for $n$ to evaluate the second integral above, we get
\beq\label{intermediate_induction}
I_{n+1}=\int\mathcal{D}\phi_{m_n}\mathcal{D}\psi~\delta(\phi-\psi-\alpha_{m_n} \phi_{m_n})exp\left[i\int \frac{1}{2}\phi_{m_n}(\Box-m_n^2+i\epsilon)\phi_{m_n}+\frac{1}{2}\psi\Box_{\delta_n}\psi\right],
\eeq
where 
\beq
\Box_{\delta_n}^{-1}=\sum_{k=0}^{n-1}\frac{\alpha_{m_k}^2}{\Box-m_k^2+i\epsilon}.
\eeq
Using eq. \eqref{master_term} once more, eq. \eqref{intermediate_induction} simplifies to
\beq
I_{n+1}=e^{i\int \frac{1}{2}\phi\Box_{\delta_{n+1}}\phi},
\eeq
where 
\beq
\Box_{\delta_{n+1}}^{-1}=\Box_{\delta_{n}}^{-1}+\frac{\alpha_{m_n}^2}{\Box-m_n^2+i\epsilon}=\sum_{k=0}^{n}\frac{\alpha_{m_k}^2}{\Box-m_k^2+i\epsilon}.
\eeq
The proof of \eqref{phi_integrate_out} is complete. 
\section{Casimir calculation}\label{app_casimir}

Let us start by defining $G_1$ and $G_2$ as the Green's functions in the presence of the potential $V_1$ and $V_2$, respectively. This means
\beq
(\tilde \Box_E+V_i)G_i=1,
\eeq
where throughout this appendix we are using matrix notation for simplicity.

One can directly verify that 
\bea
G_1(\kappa,X,Y)&=&G(\kappa,X-Y)-\frac{\lambda_1}{1+\lambda_1G(\kappa,0)}G(\kappa,X)G(\kappa,Y),\\
G_2(\kappa,X,Y)&=&G(\kappa,X-Y)-\frac{\lambda_2}{1+\lambda_2G(\kappa,0)}G(\kappa,X-a)G(\kappa,Y-a)
\eea  
satisfy $(\tilde \Box_E +V_i)G_i(\kappa,X,Y)=\delta(X-Y)$ using $\tilde \Box_E G(\kappa,X,Y)=\delta(X-Y)$.

Note that for two general operators $A$ and $B$
\beq
\mbox{Tr}\ln(AB)=\ln \mbox{Det}(AB)=\ln \mbox{Det}A+\ln\mbox{Det}B=\mbox{Tr}\ln A+\mbox{Tr}\ln B.
\eeq
Using the above, after a simple algebraic calculation we arrive at
\bea
\ln Z&&=-\frac{1}{2}\mbox{Tr}\ln\left(\tilde \Box_E+V_1+V_2\right)\notag\\
&&=-\frac{1}{2}\mbox{Tr}\ln\left(1-V_1G_1V_2G_2\right)+\frac{1}{2}\mbox{Tr}\ln\left(\tilde \Box_E G_1G_2\right)\label{ln_z_separated}.
\eea
$\mbox{Tr}\ln\left(\tilde \Box_E G_1G_2\right)$ does not depend on $a$ and is not relevant for our purpose. Also,
\bea
\mathcal{I}(\kappa,X,Y)&&\equiv (V_1G_1V_2G_2)(\kappa,X,Y)=\lambda_1\lambda_2 G_1(\kappa,0,a)G_2(\kappa,a,Y)\delta(X),\\
\mbox{Tr}_X \mathcal{I}&&=\lambda_1\lambda_2G_1(\kappa,0,a)G_2(\kappa,a,0),
\eea
satisfies $\mbox{Tr}_X \mathcal{I}^n=(\mbox{Tr}_X \mathcal{I})^n$. As a result, 
\beq
\mbox{Tr}_X\ln\left[1-(V_1G_1V_2G_2)(\kappa,X,Y)\right]=\ln\left[1-\lambda_1\lambda_2G_1(\kappa,0,a)G_2(\kappa,a,0)\right],
\eeq
which yields
\bea
\ln Z&&=-\frac{1}{2}\mbox{Tr}\ln(1-V_1G_1V_2G_2)\notag\\
&&=-\frac{1}{2}\mbox{Tr}_{x^0,{\bf x}_T}\ln\left[1-\lambda_1\lambda_2G_1(\kappa,0,a)G_2(\kappa,a,0)\right]\notag\\
&&=-\frac{1}{2}\frac{\beta}{2\pi}\frac{L^{d-1}}{(2\pi)^{d-1}}\int \dd p^0\dd {\bf p}_T\ln\left[1-\lambda_1\lambda_2G_1(\kappa,0,a)G_2(\kappa,a,0)\right],
\eea
where $\beta$ and $L$ are (IR) cutoffs on $x^0$ and transverse directions, respectively. The energy of the system in a unit area of transverse directions is
\beq
\mathcal{E}=\frac{1}{2}\frac{S_{d-1}}{(2\pi)^d}\int_0^{\infty}\dd\kappa~ \kappa^{d-1} \ln\left[1-\frac{\lambda_1}{1+\lambda_1G_0(\kappa,0)}\frac{\lambda_2}{1+\lambda_2G_0(\kappa,0)}G_0(\kappa,a)^2\right],
\eeq

where $S_{d-1}=\frac{2\pi^{d/2}}{\Gamma(d/2)}$ is the area of a $d$-dimensional unit sphere. As $\lambda_1,\lambda_2\rightarrow \infty$,
\beq
\mathcal{E}=\frac{1}{2}\frac{S_{d-1}}{(2\pi)^d}\int_0^{\infty}\dd\kappa ~\kappa^{d-1} \ln\left[1-\left(\frac{G_0(\kappa,a)}{G_0(\kappa,0)}\right)^2\right].
\eeq

\end{document}